\documentclass[12pt]{article}
\usepackage{epsfig}

\usepackage{amsmath,amssymb,amsfonts,cite}

\newcommand{\barray}{\begin{array}}
\newcommand{\eeqarray}{\end{eqnarray*}}
\newcommand{\beqarray}{\begin{eqnarray*}}
\newcommand{\earray}{\end{array}}
\newcommand{\bite}{\begin{itemize}}
\newcommand{\eite}{\end{itemize}}
\newcommand{\bmath}{\begin{displaymath}}
\newcommand{\emath}{\end{displaymath}}
\newcommand{\beq}{\begin{equation}}
\newcommand{\eeq}{\end{equation}}
\newcommand{\bea}{\begin{eqnarray}}
\newcommand{\eea}{\end{eqnarray}}
\newcommand{\bdm}{\begin{displaymath}}
\newcommand{\edm}{\end{displaymath}}
\newcommand{\bqa}{\begin{eqnarray}}
\newcommand{\eqa}{\end{eqnarray}}

\newcommand{\bd}{\begin{displaymath}}
\newcommand{\ed}{\end{displaymath}}
\newcommand{\bcen}{\begin{center}}
\newcommand{\ecen}{\end{center}}

\newcommand{\GOLEM}{{\textsc{GoSam}}}
\newcommand{\GOSAM}{{\textsc{GoSam}}}

\newcommand{\SHERPA}{{\textsc{SHERPA}}}

\newcommand{\QGRAF}{{\texttt{QGRAF}}}
\newcommand{\FORM}{{\texttt{FORM}}}
\newcommand{\SPINNEY}{{\texttt{spinney}}}
\newcommand{\HAGGIES}{{\texttt{haggies}}}
\newcommand{\SAMURAI}{{\textsc{SAMURAI}}}
\newcommand{\GOLEMVC}{{\texttt{Golem95C}}}
\newcommand{\PJFRY}{{\texttt{PJFRY}}}

\newcommand{\PYTHON}{{\texttt{Python}}}
\newcommand{\UFO}{{\texttt{UFO}}}
\newcommand{\LANHEP}{{LanHEP}}
\newcommand{\LOOPTOOLS}{{LoopTools}}
\newcommand{\CUTTOOLS}{{CutTools}}

\begin{document}
\title{Automation of One-Loop Calculations with
\GOSAM : 
Present Status and Future Outlook%
\footnote{Presented by G.~Ossola at the XXXV International Conference of Theoretical Physics 
``Matter to the Deepest'': Recent Developments in Physics of Fundamental Interactions, Ustro\'n 2011 }%
}

\author{G.~Cullen$^1$,
N.~Greiner$^{2,3}$,
G.~Heinrich$^2$,
G.~Luisoni$^4$, \\
P.~Mastrolia$^{2}$,
G.~Ossola$^{5,6}$,
T.~Reiter$^{2}$,
F.~Tramontano$^{7}$}

\date{}
\maketitle

{\small
\bcen
\vspace{2mm}
$^1$ School of Physics and Astronomy, \\
	The University of Edinburgh,
	Edinburgh EH9\,3JZ, UK\\
\vspace{1mm}
$^2$ Max-Planck Insitut f\"ur Physik, \\ F\"ohringer Ring, 6, D-80805 M\"unchen, Germany\\	
\vspace{1mm}
$^3$ Department of Physics, University of Illinois at Urbana-Champaign,\\ 1110 W Green Street, Urbana IL 61801, USA\\
\vspace{1mm}
$^4$ IPPP, %
	University of Durham, %
	Durham DH1\,3LE, UK\\
\vspace{1mm}
$^5$ Physics Department, New York City College of Technology,\\
	City University of New York, 300 Jay Street, Brooklyn NY 11201, USA \\
	\vspace{1mm}
$^6$ The Graduate School and University Center, The City University of New York\\
365 Fifth Avenue, New York NY 10016, USA \\
\vspace{1mm}
$^7$ Theory Group, Physics Department, CERN, 1211~Geneva 23, Switzerland
\ecen
}

\begin{abstract}
In this presentation, we describe the \GOSAM{} (Golem/Samurai)
framework for the automated computation of multi-particle scattering amplitudes at the
one-loop level.
The amplitudes are generated analytically in terms of
Feynman diagrams, and can be evaluated using either $D$-dimensional integrand reduction or
tensor decomposition. \GOSAM{} can be used to compute one-loop corrections to
Standard Model (QCD and EW) processes, and it is ready to link generic
model files for theories Beyond SM.
We show the main features of \GOSAM{} through its application
to several examples of different complexity.
\end{abstract}

\section{Introduction}

The discovery potential of the experimental programs at the LHC relies heavily on the availability of 
higher order corrections for many relevant processes~\cite{lh}. 
The searches for the Higgs boson and the compilation of related exclusion limits 
need precise calculations for Higgs' signal and background processes. 
Further, it will be very important to have precise theory predictions at hand 
in order to constrain model parameters in the event that a signal of New Physics will be detected. 
Therefore, it is of major importance to provide tools 
for next-to-leading order (NLO) predictions which are largely automated 
such that signal and background rates for a multitude of processes 
can be estimated reliably. 

Already some time ago, the idea of automating NLO calculations has been pursued with public programs 
like FeynArts\,\cite{Hahn:2000kx} and \QGRAF~\cite{Nogueira:1991ex} 
for diagram generation and FormCalc/LoopTools~\cite{Hahn:1998yk} and 
{\small GRACE}~\cite{Belanger:2003sd} for 
the automated calculation of NLO corrections, primarily in the electroweak sector.
In spite of this important progress, until the last few years we did not observe a large production of 
calculations of one-loop amplitudes involving more than four external legs.
Only very recently, conceptual and technical advances in multi-leg one-loop calculations allowed the calculation 
of six-point~\cite{Berger:2009ep,Berger:2009zg,KeithEllis:2009bu,Melnikov:2009wh,Berger:2010vm,
Bredenstein:2009aj,Bredenstein:2010rs,Bevilacqua:2009zn,Bevilacqua:2010ve,
Binoth:2009rv,Greiner:2011mp,Bevilacqua:2010qb,Bevilacqua:2011hy,Denner:2010jp,Melia:2010bm,Melia:2011dw,
Campanario:2011ud} 
and even seven-point\,\cite{Berger:2010zx,Ita:2011wn} processes, and opened the door to the possibility 
of an {\it automated} generation 
and evaluation of multi-leg one-loop amplitudes, rather than creating a collection 
of hard-coded individual processes. 


Even if excellent process-specific programs are available, like MCFM~\cite{Campbell:1999ah,Campbell:2000bg,Campbell:2011bn} and VBFNLO\,\cite{Arnold:2011wj}, nevertheless it is desirable to have flexible tools at hand such 
that, in the same fashion already available at the tree-level~\cite{Kanaki:2000ey, Stelzer:1994ta, Alwall:2011uj}, 
any process which may turn out to be important can be promptly evaluated at NLO accuracy.

Recently, we observed major advances in the direction of constructing packages 
for fully automated one-loop calculations, see e.g.~\cite{Ossola:2007ax,Mastrolia:2008jb,
vanHameren:2009dr,Bevilacqua:2010mx,Mastrolia:2010nb,Ossola:2010zz,Cullen:2010hz,Reiter:2010md,Hirschi:2011pa,Bevilacqua:2011xh}.
Reviewing all the concepts that lead to these advances is beyond the scope of this presentation\footnote{Additional information can be found in other talks presented in this conference~\cite{ustron}.}.
In the development of our computational tools, the OPP reduction technique~\cite{Ossola:2006us, Ossola:2007bb} and
generalized $D$-dimensional unitarity~\cite{Ellis:2008ir} turned out to be the 
most crucial ingredients. 

The purpose of this talk is to present the program package \GOSAM{}~\cite{release} which allows the automated calculation of 
one-loop amplitudes for multi-particle processes. 
The integrand is generated via Feynman diagrams, using \QGRAF~\cite{Nogueira:1991ex}, \FORM~\cite{Vermaseren:2000nd},
\SPINNEY~\cite{Cullen:2010jv} and \HAGGIES~\cite{Reiter:2009ts}. 
The individual program tasks are managed by python scripts. The only task required from the user is the preparation of
an ``input card'' in order to launch the generation of the source code and its compilation, without having 
to worry about internal details of the code generation.

Concerning the reduction, the program offers the possibility to use either the $D$-dimensional extension of
the OPP method, as implemented in \SAMURAI~\cite{Mastrolia:2010nb}, 
or tensor reduction as implemented in
\GOLEMVC~\cite{Binoth:2008uq,Cullen:2011kv} interfaced through
tensorial reconstruction at the integrand level~\cite{Heinrich:2010ax},
or a combination of both.

\GOSAM{} can be used to generate and evaluate one-loop corrections in both QCD and electro-weak theory. 
Beyond the Standard Model theories can be interfaced using
FeynRules~\cite{Degrande:2011ua, Christensen:2008py} or \LANHEP~\cite{Semenov:2010qt}.
The Binoth-Les Houches-interface~\cite{Binoth:2010xt} to programs providing the real radiation 
contributions is also included.
 
In the following, we will provide a brief description of the main features of the code, with particular attention to
the generation of the code and the various options to efficiently and automatically compute all rational terms. 
We will conclude the presentation with some examples of applications.

\section{Main features of GoSam}

\GOSAM{} produces in a fully automated way all the code required to perform the calculation of 
one-loop amplitudes, by processing the information contained in an ``input card'' prepared by the user.
The main steps in this process are: the generation of contributing diagrams, the optimization
and algebraic manipulation to simplify their expressions, and the writing of a FORTRAN code ready to be
used within a phase-space integration. The reduction of unintegrated amplitudes to linear combinations of scalar (master) integrals 
is fully embedded in the process. 

In this section, we give a brief overview of some general operations performed by \GOSAM{}. 
A complete description of the framework, together with a detailed explanation of all features available in \GOSAM{},
can be found in Ref.~\cite{release}. 

\subsection{Diagram Generation}

For the diagram generation both at tree level and one-loop level
we employ the program \QGRAF~\cite{Nogueira:1991ex}. This program
already offers several ways of excluding unwanted diagrams for example
by requesting a certain number of propagators or vertices of a certain type
or by specifying topological properties such as the presence of tadpoles or
on-shell propagators.
Although \QGRAF{} is a very reliable and fast generator we added another filter over diagrams
by means of \PYTHON{}. This gives several advantages:
first of all, the possibilities offered by \QGRAF{} are not always sufficient
to distinguish certain classes of diagrams; secondly, \QGRAF{} cannot handle the sign for 
diagrams with Majorana fermions in a reliable way; and finally, in order to fully optimize 
the reduction, we want to classify and group diagrams 
according to the sets of their propagators.

In our framework, \QGRAF{} generates three sets
of output files: an expression for each diagram for
\FORM~\cite{Vermaseren:2000nd}, \PYTHON{} code for drawing each diagram
and \PYTHON{} code for computing the properties of the diagram. 
The model information
for \QGRAF{} is either read from the built-in Standard Model file or
is generated from a user defined \LANHEP~\cite{Semenov:2010qt}
or Universal FeynRules Output (\UFO)~\cite{Degrande:2011ua} file.
The \PYTHON{} program automatically performs several operations:
diagrams whose color factor turns out to be zero are dropped;
the number of propagators containing the loop momentum, the tensor rank 
and the kinematic invariants of the associated loop integral are computed;
diagrams with a vanishing loop integral associated are detected and flagged for the diagram selection;
all propagators and vertices are classified for the diagram selection;
diagrams containing massive quark self-energy insertions or
closed massless quark loops are specially flagged.

Partitioning diagrams with similar structures and tracking their rank are very important operations
in order to reduce the number of operations performed by the reduction and allow
allow for a big gain in efficiency: after 
carrying out the tensor reduction for one diagram, all other
diagrams that contain only a subset of the denominators are reduced with virtually no additional
computational cost. This is true both in the OPP method~\cite{Ossola:2006us}
as implemented in \CUTTOOLS~\cite{Ossola:2007ax} and
\SAMURAI~\cite{Mastrolia:2010nb}
and in classical tensor reduction methods as implemented
in \GOLEMVC~\cite{Binoth:2008uq,Cullen:2011kv}, \PJFRY~\cite{Fleischer:2010sq, talkYundin}
and \LOOPTOOLS~\cite{Hahn:1998yk,vanOldenborgh:1989wn}.

During this phase, \GOSAM{} also generates a \LaTeX{} file with the drawings of all
contributing diagrams. To achieve this task, we use our own
implementation of the algorithms described in Ref.~\cite{Ohl:1995kr} and 
Axodraw~\cite{Vermaseren:1994je} to actually draw the diagrams.

\subsection{Lorentz Algebra}
Concerning the algebraic operations performed by \GOLEM{} 
to render the integral suitable for efficient numerical evaluation,
one of the primary goals is to split the $(4-2\varepsilon)$ dimensional
algebra into strictly four-dimensional objects and symbols representing
the higher-dimensional remainder.
In \GOLEM{} we have implemented the 't~Hooft-Veltman scheme (tHV) and
dimensional reduction~(DRED). In both schemes all external vectors
(momenta and polarisation vectors) are kept in four dimensions.
Internal vectors, however, are kept in the $n$-dimensional vector space.
We adopt the conventions used in~\cite{Cullen:2010jv}, where
$\hat{k}$ denotes the four dimensional projection of an in general
$n$~dimensional vector $k$. The $(n-4)$~dimensional orthogonal projection
is denoted as~$\tilde{k}$. For the integration momentum $q$ we introduce
in addition the symbol $\mu^2=-\tilde{q}^2$, such that
\begin{equation}
q^2=\hat{q}^2+\tilde{q}^2=\hat{q}^2-\mu^2.
\end{equation}
We also introduce suitable projectors by splitting the metric tensor
\begin{equation}
g^{\mu\nu}=\hat{g}^{\mu\nu}+\tilde{g}^{\mu\nu},\quad%
\hat{g}^{\mu\nu}\tilde{g}_{\nu\rho}=0,\quad%
\hat{g}^\mu_\mu=4,\quad\tilde{g}^\mu_\mu=n-4.
\end{equation}

\GOLEM{} contains a library of representations of wave functions and propagators up to
spin~two. The exact form of the interaction vertices is taken from the model files.

Once all wave functions and propagators have been substituted by the
above definitions and all vertices have been replaced by their corresponding
expressions from the model file, all vector-like quantities and all
metric tensors are split into their four-dimensional and their orthogonal
part. As we use the 't~Hooft algebra, $\gamma_5$ is defined as a purely
four-dimensional object, $\gamma_5=i\epsilon_{\mu\nu\rho\sigma}%
\hat{\gamma}^\mu\hat{\gamma}^\nu\hat{\gamma}^\rho\hat{\gamma}^\sigma$.
By applying the usual anti-commutation relation for Dirac matrices we can
separate the four-dimensional and $(n-4)$-dimensional parts of
Dirac traces.

While the $(n-4)$-dimensional traces are reduced completely to
products of $(n-4)$-dimensional metric tensors $\tilde{g}^{\mu\nu}$,
the four-dimensional part is treated such that the number of terms in the
resulting expression is kept as small as possible. Any spinor line or trace
is broken up at any position where a light-like vector appears. Furthermore,
Chisholm identities are used to resolve Lorentz contractions between
both Dirac traces and open spinor lines. If any traces remain we use
the built-in trace algorithm of \FORM{}~\cite{Vermaseren:2000nd}.

\subsection{Treatment of $R_2$ terms}
In the numerator of a one-loop diagram, terms
containing the symbols $\mu^2$ or $\varepsilon$ can lead to a so-called
$R_2$ term~\cite{Ossola:2008xq}. 
Therefore the numerator function can be written as,
\beq
\label{eq:r2}
\mathcal{N}(\hat{q},\mu^2,\varepsilon)=
\mathcal{N}_0(\hat{q},\mu^2)+
\varepsilon\mathcal{N}_1(\hat{q},\mu^2)+
\varepsilon^2\mathcal{N}_2(\hat{q},\mu^2) \, .
\eeq
It is useful to observe that the terms $\mathcal{N}_1$ and $\mathcal{N}_2$ in Eq.~\eqref{eq:r2}
do not arise in DRED, where only terms containing $\mu^2$ contribute to $R_2$.
Instead of relying on the construction of $R_2$ from specialized Feynman
rules~\cite{Draggiotis:2009yb,Garzelli:2009is,Garzelli:2010qm,%
Garzelli:2010fq}, we can generate the $R_2$ part along with all other
contribution using automated algebraic manipulations.

The code offers the option between the \emph{implicit} and \emph{explicit} construction of the $R_2$ terms. 
The implicit
construction uses the splitting of Eq.~\eqref{eq:r2} and treats
all numerator functions~$\mathcal{N}_i$ on equal grounds.
Each term  in Eq.~\eqref{eq:r2} is reduced separately and the results are added up taking into
account the powers of $\varepsilon$. 
The explicit construction of $R_2$
is based on the fact that the non purely 4-dimensional part of the
numerator function contains powers of $\mu^2$ or $\varepsilon$, and the expressions for the 
corresponding integrals are relatively simple and known explicitly.
Therefore, after separating it using the algebraic manipulation described before, 
the $(n-4)$~dimensional part is computed analytically whereas
the purely four-dimensional part is passed to the numerical
reduction. This approach also allows for an efficient calculation of the $R_2$ alone.
\subsection{Reduction to scalar (master) integrals}

\GOSAM{} allows to choose at run-time (namely without regenerating the code) the preferred method of reduction.    
Available options include the integral-level $D$-dimensional reduction, 
as implemented in \SAMURAI, 
or traditional tensor reduction as implemented in
\GOLEMVC{} interfaced through tensorial reconstruction at the integrand level,
or a combination of both.

Concerning the scalar (tensorial) integrals~\cite{thv,pv}, \GOSAM{} allows to choose among a variety of options, including
QCDLoop~\cite{Ellis:2007qk, vanOldenborgh:1990yc}, OneLoop~\cite{vanHameren:2010cp}, \GOLEMVC~\cite{Binoth:2008uq,Cullen:2011kv}, plus the recently added \PJFRY~\cite{Fleischer:2010sq, talkYundin} and \LOOPTOOLS~\cite{Hahn:1998yk,vanOldenborgh:1989wn}.
Among these codes, OneLoop and \GOLEMVC{} also fully support complex masses.

For details about the reduction methods, we refer the reader to previous presentations~\cite{Ossola:2010zz,Cullen:2010hz,Reiter:2010md} or the original articles.

\section{Examples}

The \GOSAM{} codes have been tested on several processes, starting with QCD $2 \to 2$ NLO amplitudes, but also on more challenging $2 \to 4$ (not counting decays) in the final state. Some examples are depicted in Table~1.
The full list of processes, with the details of all comparisons performed, 
is given in Ref.~\cite{release}.
\begin{table}[h]
\bcen
\begin{tabular}{|l|l|}
\hline
Process & Checked with Ref.\\
\hline
$u\overline{d}\to e^-\overline{\nu}_e\,g$&\cite{Hirschi:2011pa}\\
$e^+e^-\to e^+e^-\gamma$ {\small (QED)}&\cite{Actis:2009uq}\\
$pp \to H\,t\overline{t}$&\cite{Hirschi:2011pa}\\
$pp\to W^+W^+jj$&\cite{Melia:2010bm}\\
$pp\to b\overline{b} b\overline{b}$ &\cite{Binoth:2009rv,Greiner:2011mp,vanHameren:2009dr}\\
$pp\to W^+W^- b\overline{b}$ &\cite{vanHameren:2009dr,Hirschi:2011pa}\\
$u\overline{u} \to t\overline{t}b\overline{b}$&\cite{vanHameren:2009dr,Hirschi:2011pa}\\
$gg \to t\overline{t}b\overline{b}$&\cite{vanHameren:2009dr,Hirschi:2011pa}\\
$u\overline{d} \to W^+ ggg$&\cite{vanHameren:2009dr}\\
\hline
\end{tabular}
\ecen
\caption{Some of the processes computed and checked with \GOSAM{}}
\end{table}
\subsection{BLHA interface, \GOSAM{}, and \SHERPA}
The BLHA interface allows to link \GOSAM{} to a general Monte Carlo event generator, which is responsible for supplying the missing ingredients for a complete NLO calculation of a physical cross section. Among those, \SHERPA{}~\cite{Gleisberg:2008ta} offers the possibility to compute the LO cross section and the real corrections with both the subtraction terms and the corresponding integrated counterparts~\cite{Krauss:2001iv,Gleisberg:2007md,Schonherr:2008av}. Using the BLHA interface, we linked \GOLEM{} with \SHERPA{} to compute physical cross section for $W^{\pm}+1$-jet at NLO.

We tested our results producing distributions for inclusive and exclusive $p_{\perp}$ and $\eta$ of the jet, $H_{T}$, and for $p_{\perp}$ and $\eta$ of the leptons (details can be found in Ref.~\cite{release}). All distributions are in agreement with the ones produced using \SHERPA{} in combination with MCFM.
\begin{figure}[h]
\centering
\includegraphics[height=3.0in]{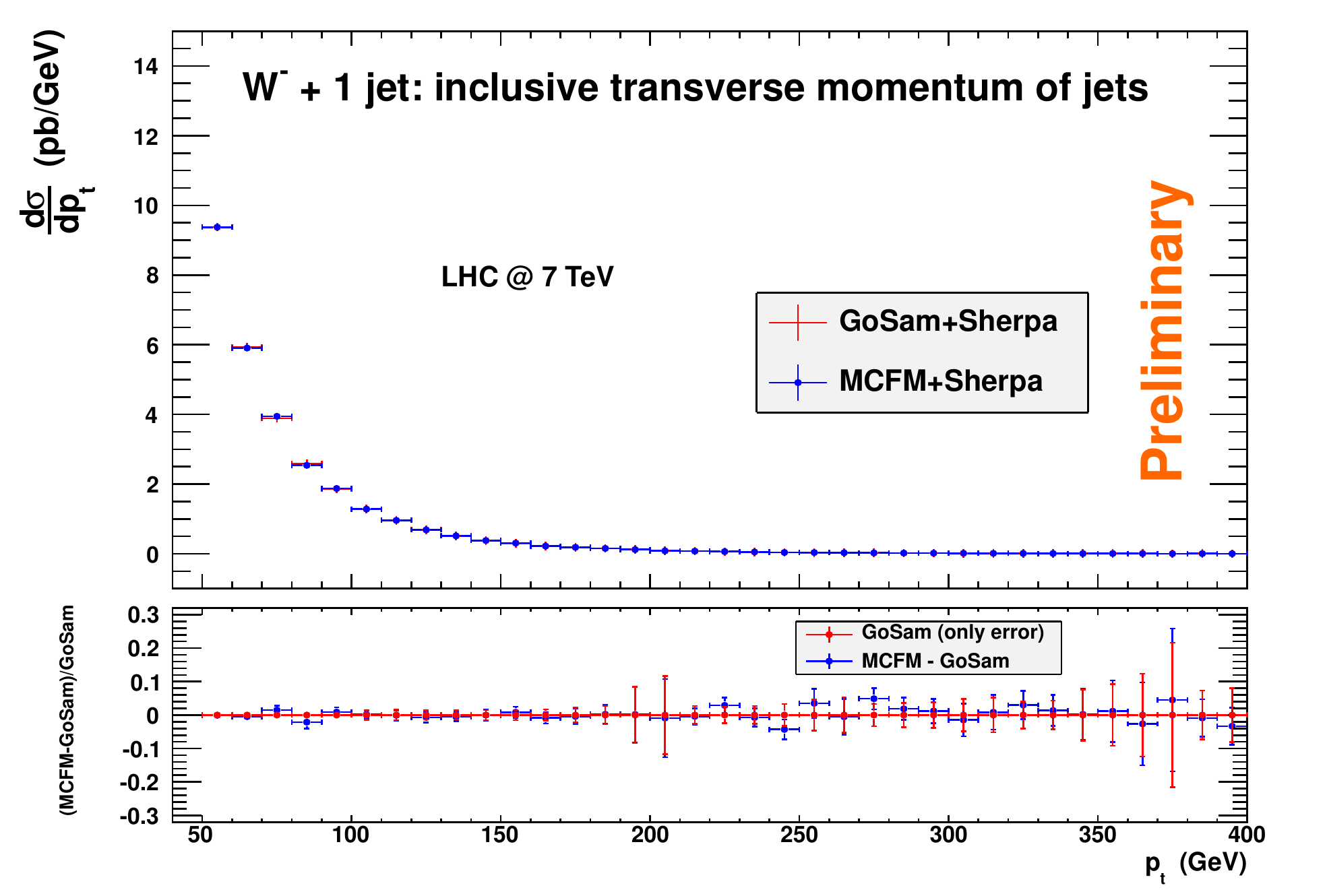}
\caption
{Comparison of the inclusive transverse momentum of $W^{-}+1$ jet between Sherpa interfaced with GoSam and Sherpa interfaced with MCFM. For comparison purpose we chose the $k_{T}$-algorithm with $p_{t,min}=50$ GeV. The bars indicate the statistical Monte Carlo error.}
\label{gs}
\end{figure}

\subsection{\GOSAM{} and Neutralino Pair Production}
As an example of the usage of \GOSAM{} with a model file different from the 
Standard Model, we calculated the QCD corrections to neutralino pair production
in the MSSM. A calculation of the total cross sections for neutralino pair production at the LHC 
is also presented in Ref.~\cite{Beenakker:1999xh}. 
The model file has been imported via the \UFO{} interface. To import such files within the \GOSAM{} setup, 
all the user has to do is to give the path to the corresponding model file in the input card.

In this example, we combined the one-loop amplitude with the 
real radiation corrections to obtain results for differential cross sections.
For the infrared subtraction terms we employed {\tt MadDipole}
\cite{Frederix:2008hu,Frederix:2010cj}, while the real emission part is 
calculated using MadGraph/MadEvent \cite{Alwall:2007st}.
The virtual matrix element is renormalized in the $\overline{MS}$ scheme, 
while massive particles are treated in the on-shell scheme. 
The renormalization terms specific to the massive MSSM particles have been added manually.
For the SUSY parameters we use the modified benchmarks point SPS1amod suggested in
\cite{Feigl:2011sw}, and use $\sqrt{s}=7$\,TeV.

\begin{figure}[h]
\centering
\includegraphics[height=3.0in]{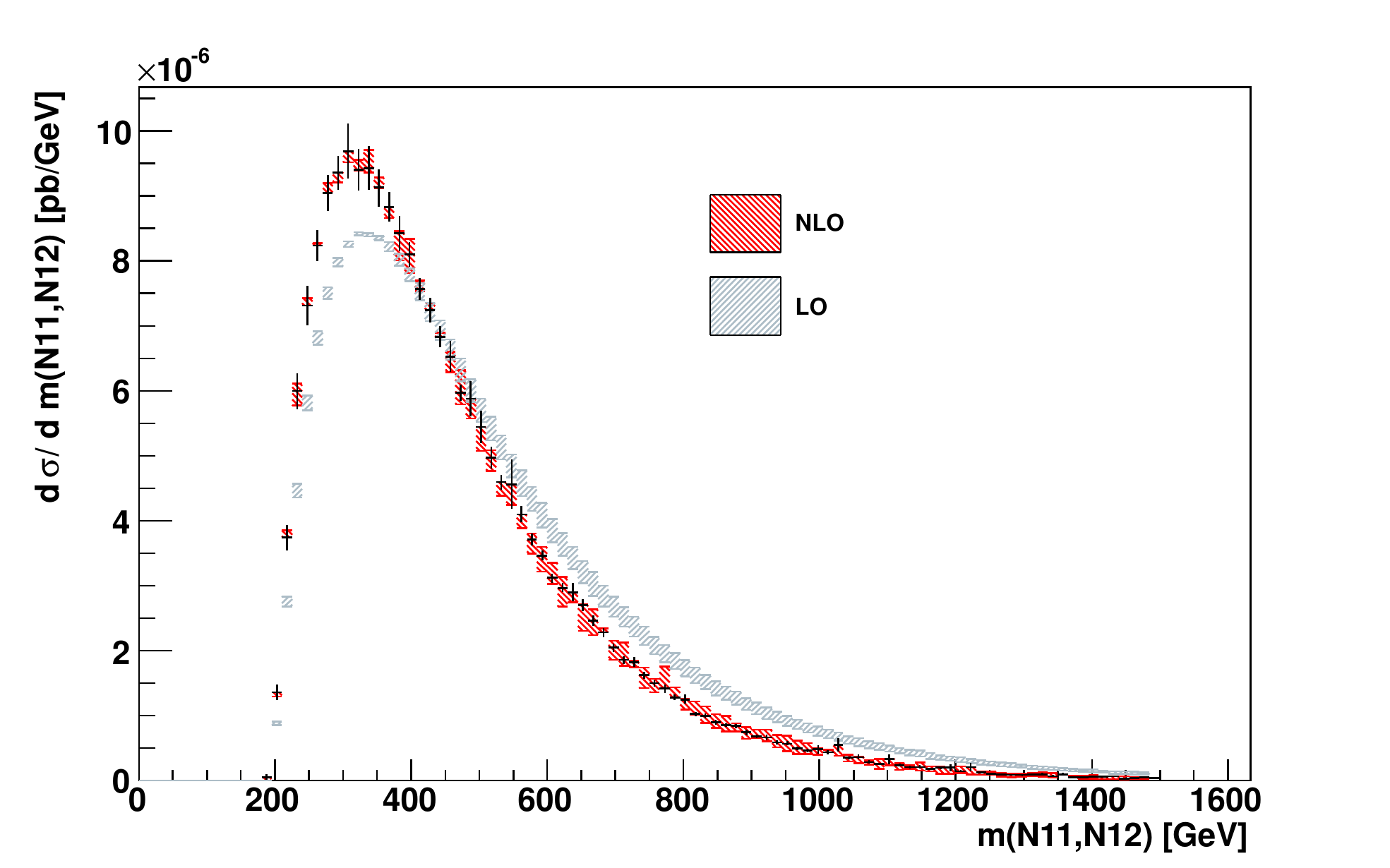}
\caption
{Comparison of the NLO and LO $m_{\chi_{1}^{0} \chi_{1}^{0}}$ distributions for the process $pp \rightarrow \chi_{1}^{0} \chi_{1}^{0}$
with a jet veto on jets with $p^{jet}_{T} > 20$ GeV and $\eta<4.5$.
The band gives the dependence of the result on $\mu = \mu_{F} = \mu_{R}$ between
$\mu_{0}/2$ and $2 \mu_{0}$. We choose $\mu_{0} = m_{Z}$. The black line gives the bin
error for the value at the central scale.}
\label{fig:qqNNm12JVfull}
\end{figure}

In Fig.\ref{fig:qqNNm12JVfull} we show the differential cross section for the 
$m_{\chi_{1}^{0} \chi_{1}^{0}}$ invariant  mass,
where we employed a jet veto to suppress large contributions from 
the channel $qg \rightarrow \chi_{1}^{0} \chi_{1}^{0} q$ which opens up 
at order $\alpha^2 \alpha_s$, but for large $p_{T}^{jet}$ belongs to
the distinct process of neutralino pair
plus one hard jet production at leading order.
\section{Outlook and Conclusions}
In the last five years, we observed major advances in our understanding of one-loop scattering amplitudes. Aside from improvement on standard tensorial techniques, the development of unitarity-based approaches, paired with the decomposition at the integrand level contained in OPP method, changed the landscape of this field, favoring the calculation of NLO amplitudes for several challenging processes and the development of new theoretical frameworks and tools for such calculations. 

For quite a long time, tree-level calculation have been fully automated and included in flexible multi-process tools~\cite{Kanaki:2000ey, Stelzer:1994ta}. The level of automation achieved by one-loop calculations is suggesting the possibility of a similar success also for the NLO.
One of the natural hopes for the future is to devise ways of extending and generalizing what we understood so far about one-loop amplitudes, in order to develop tools and methods for higher-order calculations~\cite{Gluza:2010ws,Mastrolia:2011pr,Kosower:2011ty}, but plenty of work is still needed to achieve this goal.

In this presentation, we illustrated the main features of \GOSAM{}, a new program package for the fully automated evaluation of one-loop scattering amplitudes in any renormalizable quantum field theory. In its present form, \GOSAM{} can be used to calculate one-loop corrections both in QCD and electro-weak theory and offers the flexibility to link general model files for theories Beyond the Standard Model. The amplitudes are generated in terms of Feynman diagrams and the reduction to master (scalar) integrals can be performed in several ways, which can be selected at run-time. 
 
We presented several examples of one-loop calculations performed within the \GOSAM{} framework, as well as preliminary results of interesting applications such as the interface with \SHERPA. These examples demonstrate the great flexibility, together with a competitive timing, of \GOSAM{}. We are looking forward to tackle more challenging calculations and interfacing with other existing tools in the coming months.

\vspace{3.0mm}

\noindent {\bf Acknowledgments} \\ 
G.C. and G.L. are supported by the British Science and Technology Facilities Council (STFC).
N.G. was supported in part by the U.S. Department of Energy under contract No. DE-FG02-91ER40677. 
P.M. and T.R. were supported by the Alexander von Humboldt
Foundation, in the framework of the Sofja Kovaleskaja Award
Project ``Advanced Mathematical Methods for Particle
Physics'', endowed by the German Federal Ministry of
Education and Research.
The work of G.O. was supported in part by the National Science Foundation 
under Grant PHY-0855489 and PSC-CUNY Award \#63275-00 41. The research of F.T. is supported by Marie-Curie-IEF,
project: "SAMURAI-Apps".


\end{document}